\documentclass{jetpl}
\usepackage{graphicx}
\usepackage{amssymb}
\usepackage{amsmath}
\usepackage{bm}
\usepackage{pstricks}
\usepackage{slashed}
\twocolumn
\usepackage[russian]{babel}
\lat

\title{Bilayer, hydrogenated and fluorinated graphene:  QED vs SU(2) QCD theory}

\rtitle{Bilayer, hydrogenated and fluorinated graphene:  QED vs SU(2) QCD theory}

\sodtitle{Bilayer, hydrogenated and fluorinated graphene:  QED vs SU(2) QCD theory}
\author{V. Yu. Irkhin
\/\thanks{e-mail: valentin.irkhin@imp.uran.ru} and Yu. N. Skryabin}

\rauthor{V. Yu. Irkhin, Yu. N. Skryabin}

\sodauthor{Irkhin, Skryabin}

\address{M. N. Mikheev Institute of Metal Physics, 620108 Ekaterinburg, Russia}

\abstract{
Motivated by recent experimental and calculational investigations of bilayer, hydrogenated and fluorinated graphene, we apply the formalisms of  U(1) QED (quantum electrodynamics) and SU(2) QCD (quantum chromodynamics) theories of strongly correlated state. Unlike non-bipartite triangular lattice, on bipartite honeycomb lattice
there always exists a monopole that transforms trivially under all the microscopic symmetries,  destabilizing the Dirac spin liquid (DSL), so that one can continuously tune the DSL state to the state with parent SU(2) instead of U(1) gauge group. The SU(2) theory describes a spin-liquid state which is different from usual DSL and is probably unstable with respect to Neel or valence-bond solid (VBS) phases, except for the quantum critical point. This point of view means a possibility of  VBS states in  graphene systems.
}


\begin{document}

\maketitle

\textbf{1. Introduction.}
The electron spectrum of standard graphene with weakly correlated $sp$-orbitals is described in terms of Dirac fermions corresponding to one-electron band cones with a gap which occurs owing to spin-orbit interaction. Thus the system has properties of a topological insulator, and the corresponding two-level Hamiltonian describes the anomalous  Hall effect \cite{Haldane,Hasan}.
In a strained graphene periodic gauge (pseudo-magnetic) fields with high symmetry confine the massive Dirac electrons into circularly localized pseudo-Landau levels, which can be important  for quantum valley Hall effects and quantum anomalous Hall effects \cite{K0,K1}.

In some cases graphene systems demonstrate strong electron correlations, including twisted magic-angle bilayer system where  correlated Mott state is supposed \cite{moire} and monolayer  graphene intercalated by gadolinium \cite{Gd}.
In bilayer graphene, changes in the degeneracy of the Landau levels occur at fillings corresponding to an integer number of electrons per moire unit cell.
Although the usual integer Hall effect connected with the topological Chern numbers was observed \cite{Lu}, manifestations of the correlation Hubbard subbands were found for some integer band fillings \cite{Wong}. Formation of the Hubbard splitting in such systems
can be  related to topological effects and gauge field  \cite{Scr3}.
The  Hubbard systems are similar to correlated fractional quantum
Hall states which are characterized by a topological order and
quantum entanglement and require essentially many-particle
interpretation.

In the strongly correlated regime  the excitation spectrum may change drastically. At the same time, the model still includes  Dirac fermions at the nodal points. Such a spectrum occurs in the mean-field approximation corresponding to the deconfinement spinon picture  \cite{Wen}. The corresponding non-magnetic  Dirac spin liquid (DSL) \cite{Vishwanath1} is characterized by a quantum topological order. However, the stability of DSL should be further examined and is more probable in frustrated systems.
In Ref. \cite{Scr1}, the spinon picture was applied to bilayer graphene; here we investigate the corresponding models in more detail.


The problem of magnetism in graphene materials is now extensively discussed \cite{Yazyev}; as a rule, magnetic ordering is not observed in clean systems.
In Ref. \cite{K},  a frustrated ground state for  single-side hydrogenated (C$_2$H) and fluorinated (C$_2$F) graphene was predicted, which sheds light on the absence of a conventional magnetic ordering in defective graphene demonstrated in  experiments despite presence of magnetic moments \cite{Nair,Nair1}.
This suggests a highly correlated magnetic behavior at low temperatures offering the possibility of a quantum spin-liquid state.

In the present work, we apply to this problem the gauge-field  formalism of quantum electrodynamics (QED) \cite{Kogut} and chromodynamics (QCD) \cite{Kogut1,Wang,Thomson} 
and  treat the spin-liquid  state in terms of U(1) QED and parent SU(2) QCD theories.
The former theory describes deconfinement situation and Dirac spin liquid.
The latter theory includes a monopole operator which carries trivial quantum numbers \cite{Alicea} and  the Neel to valence bond solid (VBS) quantum phase transition at the quantum critical point \cite{Wang}.
Such an approach enables us to trace the hierarchy of symmetries -- from SU(2) to U(1) and Z$_2$ spin liquids, the latter being the most stable one.

\textbf{2. Formalism of deconfined quantum critical points}.
Modern theoretical understanding of the paramagnetic Mott state and continuous zero-temperature metal-insulator   transitions is
based on slave-particle theories. These include separated (deconfined)
spin and charge  degrees of freedom  of the
electron. The charge ones are assigned to a boson which is gapless and condensed in
the metal phase, but gapped and disordered in the insulator.
Thus the  transition into the metallic phase is described as a
Bose-Einstein condensation  of charged bosons coupled to a gauge
field \cite{Vojta}.

With increasing the Hubbard  $U$, there is a continuous transition form a metal to an insulator with a ghost Fermi surface of neutral fermionic excitations (spinons). Formally, the electron annihilation operator is represented as  a product of a charged boson, $b$, and a neutral spinful fermion $f_\alpha$ (the spinon), $ c_\alpha=b f_\alpha$ \cite{Sachdev4}. The superfluid phase of the bosons is actually a metallic Fermi liquid state for the physical electrons. When replacing $b$ by its $c$-number expectation value $\langle b \rangle$, the spinons acquire the same quantum numbers as the $c_\alpha$ electrons, so that the $f_\alpha$ Fermi surface describes a conventional metal. The Mott insulator for the bosons is also a Mott insulator for the electrons, with a gap to all charged excitations. However, the $f_\alpha$ Fermi surface survives in this insulator and describes a continuum of gapless  spinons.

The most complete study of such a transition has been carried out on the honeycomb graphene lattice at half-filling where the metallic state is actually a semi-metal which only contains {\it gapless} electronic excitations at isolated Fermi points. The electronic states near these points have a Dirac-like spectrum, so that  a relativistic Dirac formalism can be used \cite{75,76}.
The corresponding action 
describes a conformal field theory (CFT). Thus we have an {\it  algebraic spin liquid}
with a power-law spectrum  and no well-defined quasiparticles.
Similar algebraic spin liquids can be also considered on the square
and kagome lattices. Although the bare-lattice fermion dispersion
does not lead here to a Dirac spectrum, allowing for non-zero  fluxes on the plaquettes
the resulting flux states can acquire such a spectrum \cite{Sachdev4}.
Last time, spin-liquid states on the kagome lattice have been actively studied \cite{kag,Mo}, in particular considering exotic excitations on the dual honeycomb lattice \cite{Mo}.

The phase transitions of interest may have an enlarged emergent symmetry, which rotates the Landau order parameters. In particular, for a spin-1/2 square lattice antiferromagnet the second-order  transition between the Neel ordered state and the VBS paramagnet can be described by the ``non-compact'' (i.e., flux-conserving) CP$^1$ (NCCP$^1$) field theory  \cite{Wang}
\begin{equation}
\label{nccp1su2} {\cal L}_0 = \sum_{\alpha = 1,2} |D_b z_\alpha|^2
- \left(|z_1|^2 + |z_2|^2\right)^2.
\end{equation}
Here $z_\alpha$ ($\alpha = 1,2$) are bosonic spinons coupled to a
dynamical $U(1)$ gauge field $b$, and $D_{b,\mu}=\partial_{\mu}-ib_{\mu}$
is the covariant derivative.
This model has a global $SO(3)$ symmetry ($z_\alpha$ transforms
as a spinor). It also has a global U(1) symmetry associated with the conservation
of the flux of field $b$, which is not an exact symmetry in the microscopic lattice model. Therefore, monopole operators  picking up a phase under a U(1) rotation should be added to the Lagrangian.
A NCCP$^1$ model, which naively possesses only SO(3)$\times$O(2) symmetry, has an emergent SO(5) symmetry at the critical point,  as demonstrated numerically (see \cite{Nahum,Wang}).


The gauge theory of spin-$1/2$ systems can be also formulated in terms of the fermionic spinons  $f_{i\alpha}$ by using the decomposition
$
{\bf S}_i=\frac{1}{2}\sum f^{\dagger}_{i\alpha}\mbox {\boldmath
$\sigma $}_{\alpha\beta}f_{i\beta}$,
with $\mbox {\boldmath$\sigma $}$ the Pauli matrices.
To take into account  the constraint $\sum_{\alpha}f_{i\alpha}^{\dagger}f_{i\alpha}=1$ beyond the mean-field approach, one has to introduce a dynamical $U(1)$ gauge field $a_{\mu}$ coupled to the fermions $f$, i.e. $t_{ij} \rightarrow t_{ij}\exp(ia_{ij})$ for the hopping integrals \cite{Vishwanath1,Wen1}.
The Dirac dispersion
with four flavors of Dirac fermions ($N_f=4$, two spin and two valley labels)
can be realized on the honeycomb lattice with  nearest-neighbor hopping, and on other lattices 
with appropriate choice of hoppings $t_{ij}$.
The non-bipartite nature (second-neighbor hopping on bipartite lattice) is needed to make sure that the gauge group is $U(1)$ rather than $SU(2)$~\cite{Wen1}.
The mean-field Hamiltonian  breaks lattice symmetry, but the spin-liquid state has all the lattice symmetry after we incorporate the above constraint in terms of projective symmetry groups \cite{Wen1}.

Then in the low energy  infrared (IR) limit, the theory reduces to the Lagrangian of  Quantum Electrodynamics in 2+1 dimensions, QED$_3$ ($N_f=4$) \cite{Vishwanath}:
\begin{equation}
\label{qedL}
\mathcal{L}=i\sum_{j=1}^{4}\bar{\psi}_j\slashed{D}_a\psi_j + \frac{1}{4e^2}F^2_{\mu\nu},
\end{equation}
where $\slashed{D}_a=\gamma^{\mu}D_{a,\mu}$ is the
gauge covariant Dirac operator,  $\psi_j$ is a two-component Dirac fermion with four flavors labeled by $j$, $\bar{\psi}=\psi^{\dagger}\gamma^0$,  $a_\mu$ is a dynamical $U(1)$ gauge field, and $F_{\mu\nu}$ is the Maxwell field. One chooses $(\gamma_0, \gamma_1, \gamma_2)=(i\mu^2, \mu^3, \mu^1)$ where $\mu$ are the Pauli matrices in the Dirac space.  The theory assumes that the $U(1)$ gauge flux, i.e. the total flux of the magnetic field, $J_{\mu}=\frac{1}{2\pi}\epsilon_{\mu\nu\lambda}\partial_{\nu}a_{\lambda}$ (that corresponds to a global U(1) symmetry called U(1)$_{\rm top}$), is conserved. The conserved charge is simply the magnetic flux of the emergent U(1) gauge field. Then one can  define monopole operators which carry this global U(1)$_{\rm top}$ charge, i.e.  change total gauge flux by 2$\pi$.
This theory,  referred to as noncompact $N_f=4$ QED$_3$ theory, flows to a stable critical fixed point in the IR limit.


For bipartite lattices, in the  mean field approximation \cite{Wen1} one can  continuously tune the Hamiltonian, without breaking any symmetry or changing the low-energy Dirac dispersion, to reach a point with particle-hole symmetry,
$f_{j\alpha}\to (-1)^j i\sigma^2_{\alpha\beta}f^{\dagger}_{j\beta}$ \cite{Vishwanath}.
This theory will then have a larger gauge symmetry of SU(2)$_{\rm g}$. 
The low energy theory again has four massless Dirac cones with two valleys, each forming a fundamental under both gauge SU(2)$_{\rm g}$ and spin SU(2)$_{\rm s}$. The continuum field theory of such state, described by an SU(2) gauge field coupled to four Dirac cones, is  QCD$_3$ theory  with $N_f=2$. The Lagrangian is given by
\begin{equation}
\mathcal{L} =\sum_{v=1,2} i \bar{\psi}_v \gamma^{\mu} (\partial_{\mu} - i a_{\mu})\psi_v,
\label{1}
\end{equation}
where $a$ is an SU(2) gauge field, and $\psi_{1,2}$ are two SU(2)-fundamental fermions.
This theory can be obtained from the square lattice spin-$1/2$ model through
a $\pi$-flux mean field ansatz, and has an SO(5) symmetry
which becomes manifest when (\ref{1}) is written in terms of Majorana fermions \cite{Wang}.
Fluctuations about the $\pi$-flux state are described by $(2 + 1)$-dimensional QCD$_3$ with a SU(2) gauge group
\cite{Thomson}.

In an alternative theory,  the SU(2) gauge symmetry  in QCD$_3$ is lowered to U(1) owing to the  Higgs phenomenon \cite{Wang}:
\begin{equation}
\mathcal{L}=\sum_{i=1}^{4}i\bar{\psi}_i\gamma^{\mu}(\partial_{\mu}-ia_{\mu})\psi_i+
(\lambda\mathcal{M}_a+{\rm h.c.}),
\label{2}
\end{equation}
where $a_{\mu}$ is now a U(1) gauge field, and the term $\mathcal{M}_a$ represents
 instanton tunneling.

{The flavor symmetry of
QCD$_3$ at $N_f = 2$ is SO(5)}.
In both the theories (\ref{1}) and (\ref{2}), the Dirac fermions
transform in the spinor representation of the
SO(5) group, the SO(5)-vector operators being time-reversal invariant mass operators.
None of the duality field theories
possesses the full SO(5) symmetry
(combining antiferromagnetic and VBS order parameters) explicitly.
While the IR fates of the theories (\ref{1}) and (\ref{2}) are unknown,
both theories have {the same symmetry anomaly as the deconfined critical point}.
Therefore, it is  probable that at leat one of these theories will flow to the deconfined critical point in the IR limit.

At the lattice scale the SU(2) gauge symmetry can be lowered  to U(1) owing to the Higgs mechanism by restoring a weak hopping which breaks the  particle-hole symmetry, so that the usual U(1) Dirac spin liquid will be recovered at low energies \cite{Vishwanath}.

The behavior of SU(2) theory is not exactly known, and a number of scenarios can be proposed \cite{Wang}.
The simplest (trivial) scenario is that  $N_f = 2$ QCD$_3$ will confine, and in the
process spontaneously break SO(5) symmetry by generating a condensate and
selecting either the VBS state or the Neel state,  a quantum deconfined critical point being absent.
Second, we could in principle flow to a stable gapless fixed point at which all perturbations
which preserve lattice and SO(3) symmetries are irrelevant.
We would then have a completely stable gapless spin liquid phase with emergent SO(5) symmetry.
Finally, $N_f = 2$  QCD$_3$ could flow to a gapless fixed
point which is stable in the presence of SO(5), but
 allows a single relevant perturbation when SO(5)
is broken to the physical symmetry.
Then QCD$_3$ (tuned to an SO(5) symmetric point)
describes the deconfined critical point, and perturbing it
drives it into either the VBS phase or the Neel
phase. This is the most probable  scenario.
Note that the competition of the VBS and Neel states was treated
in large-scale quantum Monte Carlo simulations  on the honeycomb lattice with cluster charge interactions, which was proposed as an effective model for  twisted bilayer graphene near half-filling \cite{Lee}.

\textbf{3. Monopoles on bipartite and non-bipartite lattices}.
 A second-order transition between two distinct symmetry-broken phases
(forbidden by the Landau theory)  is possible provided that the  special critical excitations -- monopole (instanton) topological defects have nontrivial quantum numbers, i.e., skyrmion defects (which carry quantum numbers under lattice symmetries)  occur in the Neel phase, and vortices (which have spin 1/2) in the VBS phase.
A similar situation takes place in Weng's treatment of competition of superconducting and antiferromagnetic phases with participation of vortices around spinons and holons \cite{Weng}.

According to Weng's theory,
the antiferromagnetic and  superconducting phases are dual: in the former phase, holons are confined while spinons are deconfined and condensed, and vice versa in the latter phase \cite{Weng}.
Another example is provided by a direct transition between a Neel ordered Mott insulator and a two-sublattice $d_{x^2-y^2}$ superconductor \cite{Vishwanath3}.
Here vortices of the AFM
are charged and the
vortices of the superconductor carry spin. The condensation of either
type of vortices drives the system between the two phases.
The topological defects of these two
 ``conventional'' phases carry unconventional quantum
numbers since both phases are closely related
to a topological band insulator which in fact has a short-range entanglement rather than topological order \cite{041004}.
%
{An important theoretical feature is that
there is no emergent U(1) symmetry at criticality.}

The Dirac spin liquid can be unstable with respect to proliferation of monopoles, and different ordered states
 can be reached from DSL \cite{Vishwanath1}.
Depending on the symmetry of the interaction, a mass term  can be generated.
This is described  by the  Gross-Neveu type model:
 \begin{equation}
\mathcal{L}=\sum_{i=1}^{4}\bar{\psi}_ii\slashed{D}_a\psi_i + g \bm \phi \cdot \bar \psi {\bm M} \psi + (\partial_\mu \bm \phi)^2 - u \bm \phi^2 - \lambda \bm \phi^4.
\end{equation}
Here $\bm \phi$ are bosonic fields which can  be of a scalar  or vector type depending on the type of generated mass $\bar \psi \bm M \psi$,  $\bm M$ being either the identity or a vector such as $\bm M = (M_{01}, M_{02}, M_{03})$.



The symmetry properties of the magnetic monopoles are  different on different lattices \cite{Vishwanath1}.
The difference is owing to that on bipartite lattices
one can continuously tune the DSL state to another spin-liquid state with SU(2) (instead of U(1)) gauge group.

%
For bipartite (honeycomb and square) lattices, there is always one monopole operator
which transforms trivially under all microscopic symmetries
owing to the existence of a parent SU(2) gauge theory \cite{Alicea}.
For the honeycomb lattice, it is ${\rm Re} \Phi_3$ \cite{Vishwanath1}.
This is a spin singlet which carries no non-trivial quantum numbers and therefore
provides an allowed perturbation to the Hamiltonian,  destabilizing  DSL.
On the non-bipartite (triangular and kagome) lattices such a destabilization does not occur.


\textbf{4. Discussion}.
We have seen that the situation for bipartite (honeycomb) and non-bipartite (triangle) lattices is different.
For bipartite situation,  {there is no additional
topological symmetry since the flux of SU(2) gauge field  (unlike that of U(1) field)
is not conserved   \cite{Vishwanath}.} Thus  a non-trivial topology is absent.
For the non-bipartite lattice,  monopoles do not prevent stability of spin liquid (DSL is transformed to Z$_2$ spin liquid by inclusion of the Higgs field).
For frustrated bipartite lattices, spin liquid is expected to exist at the quantum critical point only, but the quantum critical behavior can be observed at finite temperatures.

The conditions of a spin liquid formation in the Heisenberg model are rather strict even for the triangular lattice  \cite{Vishwanath}.
In the doped case frustrations owing to current carriers can play a role \cite{Wen}.
 In the triangular lattice Hubbard model, a spin liquid state can occur at intermediate values of the Hubbard $U$ with the transition to the ordered Neel state at larger values  \cite{22}.
For the honeycomb lattice Hubbard model,  a spin liquid phase may also occur in some
range of $U$  \cite{24}.

As discussed in Ref.\cite{Scheurer}, experimental data for twisted bilayer graphene indicate that the electron charge density is concentrated on a moire triangular lattice, so that the consequences of local correlations should be similar to those on the triangular lattice. On the other hand, symmetry and topological aspects of the band structure require that the model should be formulated using the Wannier orbitals of a honeycomb lattice.
Taking into account  momentum-dependent form factors in the magnetic moments, different models of triangular-symmetry antiferromagnetism  in  bilayer graphene were treated  in Ref.\cite{Scheurer}. Besides the minimal phenomenological model on the triangular lattice, the authors considered the model where the spin density is centered on the bonds of the dual bipartite honeycomb lattice.
The half-filled triangular lattice model and the quarter-filled honeycomb-lattice model can be consistent with experimental observations.
The half-filled honeycomb-lattice model requires the additional
Kekule VBS order which is in agreement with the Monte Carlo calculations \cite{Lee}.
The results of Ref.\cite{Scheurer} can also be extended to the case where the antiferromagnetic order is not long-ranged, but demonstrates quantum fluctuations in a state with Z$_2$ (toric code) topological order including spinons. Note that a spinon picture can be formulated in the case of  the finite-$U$ Hubbard model  \cite{Punk}.

Formation of a gapless RVB state on the anisotropic kagome  lattice  (having dual honeycomb lattice) with application to the system LiZn$_2$Mo$_3$O$_8$ was considered in Ref.\cite{Mo}.

Triangular versus honeycomb lattice problem for bilayer graphene was considered in \cite{Po}. Although the charge density is concentrated on the triangular lattice sites of the moire  pattern, the Wannier states of the tight-binding model must be centered on different sites which form a honeycomb lattice.
A simple Anderson's RVB picture of quantum spin liquids with neutral spinons and bosonic holons was also discussed  in Ref.\cite{Po}.

Generalized triangular lattice Hubbard models have been proposed
to describe  flat moire bands  in twisted van der Waals  transition metal dichalcogenide heterobilayers \cite{Tutuc}.
Recently a heterostructure of ABC-stacked trilayer graphene and boron nitride, which also forms a triangular moire superlattice even at zero twist angle, was studied \cite{boron}. A possibility of a  fractional quantum anomalous Hall effect in twisted bilayer graphene aligned with its hexagonal boron nitride substrate was considered in Ref.\cite{Senthil}.

An effective Heisenberg model was built in Ref. \cite{K} for the  C$_2$H and and C$_2$F systems, which includes competing exchange interactions on different $p$-orbitals and combines features of honeycomb and triangle lattices. The presence of antiferromagnetic interactions on the triangular lattice of the moments leads to the instability of the collinear magnetic ordering due to frustration. The case of C$_2$H turns out to be even more complicated due to the presence of the two nonequivalent magnetic sublattices comprising the honeycomb lattice. According to the calculation  \cite{K}, a frustrated model with triangle features can be applied. Thus frustration can lead to the DSL state since monopoles are irrelevant.
On the other hand, we can propose existence of the dual VBS state in hydrogenated and fluorinated graphene with sublattice-disordered  occupations.

The authors are grateful to M.I. Katsnelson and S.V. Streltsov for helpful discussions.
The research was carried out within the State Assignment
of Education Ministry  of Russia
(theme ``Flux'' No AAAA-A18-118020190112-8 and theme ``Quantum'' No. AAAA-A18-118020190095-4).

\end{document}